\documentstyle[12pt]{article}

\begin{document}
\newcommand{\beq}{\begin{equation}}
\newcommand{\eeq}{\end{equation}}
\newcommand{\BoldM}[1]{\mbox{\boldmath{${#1}$}}}
\newcommand{\ROSAT}{{\it ROSAT}}
\newcommand{\ASCA}{{\it ASCA}}
\newcommand{\const}{{\it const}}
\newcommand{\dg}{^{\rm o}}
\renewcommand{\thefootnote}{\fnsymbol{footnote}}
\def\spose#1{\hbox to 0pt{#1\hss}}
\def\lta{\mathrel{\spose{\lower 3pt\hbox{$\mathchar"218$}}
     \raise 2.0pt\hbox{$\mathchar"13C$}}}
\def\gta{\mathrel{\spose{\lower 3pt\hbox{$\mathchar"218$}}
     \raise 2.0pt\hbox{$\mathchar"13E$}}}
\parskip=0pt
\parindent=0pt
\normalsize
\pagestyle{empty}
\par\noindent
\centerline{\large\bf Doppler Tomography of Relativistic Accretion Disks}
\bigskip\bigskip\bigskip\bigskip\bigskip

\noindent
\centerline{\large Vladim\'{\i}r {\sc Karas}\footnotemark{}
  and Pavel {\sc Kraus}}

\footnotetext{Also at
 Scuola Internazionale Superiore di Studi Avanzati, Trieste;
 Department of Astronomy and Astrophysics, G\"oteborg University and
 Chalmers University of Technology, G\"oteborg}
\bigskip\bigskip\bigskip\bigskip

\centerline{\it Astronomical Institute, Charles University,
 \v{S}v\'edsk\'a 8,}
\centerline{\it CZ-150 00 Prague, Czech Republic}
\vspace*{5cm}

\centerline{Scheduled for}
\centerline{Publications of the Astronomical Society of Japan,}
\centerline{ Vol. 48 (October 1996)}

\parindent=25pt
\parskip=3pt
\newpage
\pagestyle{plain}
\pagenumbering{arabic}
\noindent
\centerline{\large\bf Abstract}
\medskip\par
Spectral lines from a source orbiting around a compact object are
studied. Time variations of observed frequency and count rate
due to motion
of the source and gravitational lensing are considered.
Gravitational field of the central object is described by the Kerr metric.
It is shown that: (i)~simultaneous temporal and frequency resolution enables
us to restrict parameters of the model (inclination angle,
position of the source, angular momentum of the black hole);
(ii)~techniques of image restoration, familiar from other fields of
astronomy, can be applied to study inner regions of active
galactic nuclei. This contribution is relevant for extremely variable
X-ray sources with high parameter of efficiency, such as
Seyfert~1 galaxy PHL~1092 observed by \ROSAT.

\bigskip\bigskip\par

\noindent
{\large\bf 1.\quad Introduction}
\medskip\par
Tomography is a method of image restoration which has been successfully
applied in various fields, particularly in medicine. Image reconstruction,
on the other hand,
is one of the typical problems in astronomy where the situation is
rather specific because astronomical objects are very distant and can
be viewed from a single direction only. Rotation of the source is thus
essential, and one needs to understand how rotational motion is linked
to observed profiles of spectral lines. Vogt, Penrod \& Hatzes (1987)
describe the tomographical method
for mapping the surface of rotating stars. It is assumed that irregularities
(spots) on the stellar surface radiate in a spectral line, observed
frequency of which is periodically affected by the Doppler effect.
An analogous approach was applied in order to study accretion disks in
cataclysmic variable stars (Marsh, Horne 1988; Kaitchuck et al.\ 1994)
and Algol-type binaries (Albright, Richards 1996).
It has been demonstrated that Doppler
tomography can be very useful for tracing the structure of the disk surface
when it is seen at a sufficiently large inclination.
Similar techniques have been used recently to study detached binaries and
multiple stars (Simon, Sturm 1994; Hadrava 1995). All these methods are based
on the fact that the line spectrum of the source is affected by the radial
velocity of its individual
components with respect to the observer. Knowing the composite
spectrum at different phases, the problem of the line formation can be inverted
and contributions from separate components of the source can be revealed.

The present paper deals with two topics: (i) the applicability of Doppler
tomography
to study rotation and surface structure of disks around compact objects
is discussed, and (ii) the relevance of time-resolved spectral-line profiles
for study of the rapid X-ray variability of active galactic nuclei (AGN)
is explored.
(As for Doppler tomography, we retain the original terminology,
although it is less appropriate in the
context of general relativity in which the distinction between
Doppler shift and the gravitational shift of the photon
energy cannot be defined unambiguously.) The motivation for
investigating these problems comes from a widely accepted model of
AGNs with a central black hole surrounded by
an accretion disk (see, e.g., Rees 1984).
It has been realized by several authors that double-peaked asymmetric
profiles of the H$\alpha$ line observed in some galactic nuclei
can be explained by relativistic effects (Chen, Halpern, Filippenko 1989;
Eracleous, Halpern 1994).
Theoretical line profiles have been studied by numerous authors who also
discussed specific effects of general relativity: frame dragging
(Kojima 1991; Laor 1991; Bromley, Chen, Miller 1996),
multiple images (Bao, Hadrava, {\O}stgaard 1994),
and self-gravity of relativistic disks (Karas, Lanza, Vokrouhlick\'y 1995).

For a long time, some AGNs have been recognized as rapidly variable X-ray
sources (Feigelson et al.\ 1986; Mushotzky, Done, Pounds 1993;
Papadakis, Lawrence 1995). Apart from blazars, the class of I~Zw~1 objects
(Phillips 1976; Boller et al.\ 1993) has been distinguished by
the high value of their efficiency parameter:
$\eta\approx5\times10^{-43}\Delta L_{\rm X}/\Delta t\gta0.1$ (Fabian 1979).
This parameter characterizes temporal change of the source X-ray luminosity,
$\Delta L_{\rm X}$ [erg\,s$^{-1}],$ during the relevant variability
time-scale $\Delta t\approx10^2$--$10^5$ seconds.
For example, a member of the I~Zw~1 class,
narrow-line Seyfert~1 galaxy PHL\,1092 has recently been studied in detail
(Forster, Halpern 1996) and it has been speculated that the remarkably
high value of its efficiency parameter $(\eta\gta0.6)$ calls for a
non-standard model of the object (strongly anisotropic emission or
accretion onto an extremely rotating black hole; Kwan et al.\ 1995).
It is suggested in the present contribution that the lensing effect
acting on a source of radiation orbiting a black hole (as discussed by
numerous authors in another context; cf.\
Abramowicz, Bao, Lanza, Zhang 1991; Rauch, Blandford 1994)
should also be considered as a possible explanation of such extreme
variability. The model can be submitted to an observational test
by studying time-resolved spectral
lines which are formed in that region of the source responsible for
variability. Observations have not yet provided sufficient
resolution for serious spectroscopy of this type.

The most prominent relativistic effects can be expected for
spectral features around 7~keV because this complex is presumably
formed in the innermost regions of accretion disks (Fabian, George 1991;
Matt, Perola, Piro, Stella 1992). The improved technology of
\ASCA's detectors makes high-resolution spectroscopy of the
X-ray sources feasible (Serlemitsos et al.\ 1995) so that the
above-mentioned theoretical works on line profiles now become timely.
Possible future X-ray spectroscopy missions will provide temporal resolution
of the spectral lines which has not yet been much discussed in the
literature. The subsequent section of this paper deals with predicted line
profiles
as a function of time. Specific features of the frequency
shift relevant for the tomographical method of the disk mapping are mentioned.
The main aim of this discussion is to point out the differences
between accretion disks around compact objects and other situations
where the effects of relativity are negligible.
\bigskip\bigskip\bigskip\par

\noindent
{\large\bf 2.\quad Doppler Tomography of Accretion Disks}
\bigskip\bigskip\par

\noindent
{\it 2.1\quad Details of the Model}
\bigskip\par

It is assumed that a localized region of enhanced
emissivity is located on the surface of an accretion disk.
The possible existence of such irregularities
has been discussed by several authors because it has important consequences
for variability of AGNs (Abramowicz, Lanza, Spiegel, Szuszkiewicz 1992;
Sivron, Caditz, Tsuruta 1996). It is assumed that the disk resides in the
equatorial plane of a Kerr black hole.
It will be shown that the observed flux and frequency of a spectral line
as functions of time determine, to a certain degree, the
inclination of the disk
and the angular-momentum parameter of the black hole, $a$ $(0\leq a\leq1).$
The relative frequency shift $g$ of radiation (ratio of frequency
observed by a distant observer to frequency emitted in a
local frame corotating with the disk material)
can be expressed, within the approximation of geometrical
optics, in terms of four-momentum of photons, $\BoldM{p},$ and
four-velocities $\BoldM{u}_{\rm em},$ $\BoldM{u}_{\rm obs}$ of the emitting
material and of the distant observer, respectively:
\beq
g\equiv\frac{\nu_{\rm obs}}{\nu_{\rm em}}
 =\frac{\BoldM{p\cdot u}_{\rm obs}}{\BoldM{p\cdot u}_{\rm em}};
\label{g}
\eeq
$g$ is a function of polar coordinates $\{r,\phi\}$ in the disk
plane, defined in such a way that
$\{\phi=0,\theta=\theta_{\rm obs}\}$ represents a direction to the observer
($\theta_{\rm obs}=90\dg$ when the disk is seen edge-on).
In order to minimize the number of free parameters, Keplerian rotation law
is assumed: $\Omega\propto1/(r^{3/2}+a)$ in usual dimensionless
geometrized units. This restriction is in agreement with our assumption
of geometrically thin, equatorial disk. The method can be applied
to a different rotation law analogously.

Radiation flux from the source is obtained by integrating
the observed intensity $I_{\rm obs}$ over the observer's local sky.
The resulting value of the flux is determined by distribution of $g$ and
$I_{\rm em}$ (locally emitted intensity) in the disk
$(I_{\rm obs}/I_{\rm em}=g^3)$, and it also depends on the shape of
light trajectories which are
affected by the presence of the central object (lensing effect).
Various semi-analytical and completely numerical approaches have
been developed for efficient integration of the
observed radiation flux (Pineault, Roeder 1977;
Luminet 1979; Asaoka 1989). We employed the method
described by Karas, Vokrouhlick\'y \& Polnarev (1992).
\bigskip\bigskip

\noindent
{\it 2.2\quad Line Profiles Resolved in Time}
\bigskip\par

First, it is instructive to approximate light rays by straight lines because
the problem of observed line profiles can then be solved analytically
without difficulty (Gerbal, Pelat 1981; Kojima 1991).
Let us consider a point-like source moving along a circular orbit with radius
$r=r_{\rm em}$ and emitting, in its rest frame, a spectral
line with Dirac function--type profile:
\beq
I_{\rm em}(\nu,r,\phi)\propto
\delta(\nu-\nu_{\rm em})\delta(r-r_{\rm em})\delta(\phi-\Omega t).
\label{i}
\eeq
The observed flux from a single source is a simple periodic function of time,
\beq
F(\nu,t)\propto\int g^3I_{\rm em}(\nu/g,r,\phi-r\cos\phi\sin\theta_{\rm obs}),
\label{f}
\eeq
and its observed phase $\varphi$ can thus be normalized to the unit interval,
$0\leq\varphi\leq1$.
The constant of proportionality in eq.~(\ref{f}) is undetermined
unless a specific physical mechanism generating radiation is assumed,
but for the ratio of the count rates at the maximum and
the minimum observed frequency one can write
\beq
\frac{F(\nu_{\rm max})}{F(\nu_{\rm min})}=
 \frac{\nu_{\rm max}^3}{\nu_{\rm min}^3},\quad
\frac{\nu_{\rm max}}{\nu_{\rm min}}=
 \frac{r^{1/2}_{\rm em}+\sin\theta_{\rm obs}}
 {r^{1/2}_{\rm em}-\sin\theta_{\rm obs}};
 \label{ratio}
\eeq
$r_{\rm em}$ can be estimated from relation (\ref{ratio}).
Superposition of radiation from multiple sources results in a fluctuating
signal but, in principle, individual contributions can still be restored.
In practice, however, temporal and frequency resolution of
observational data is limited, and one has to work with a more
complicated local profile of the line.
An appropriate technique of image restoration has been discussed by
Marsh \& Horne (1988). These authors ignored any relativistic corrections
(frequency shift was determined solely by radial velocity of the source),
nevertheless, their approach can be applied to our situation if
eqs.~(\ref{g})--(\ref{i}) for the redshift factor and intensity are
taken into account. The difference becomes evident upon comparison
of our Fig.~\ref{fig1} with Fig.~1 of Marsh \& Horne: Notice the
asymmetry of isocontours
of $g$ with respect to the vertical axis. The shape of the contour-lines is
determined by mutual competition of the overall gravitational redshift
with the Doppler effect due to rotation (right-hand
side of the disk approaches the observer and the corresponding
observed frequency is thus
increased). As a result of this asymmetry, observed profiles of spectral lines
are also asymmetric about the emitted frequency.

The lensing effect, neglected in the discussion above, calls for a
correction of the observed flux. The magnitude of the error
introduced by this neglect was discussed by several authors
(Kojima [1991] corrects an inconsistency in the work of Gerbal \& Pelat
[1981], but, in most situations, the difference does not produce a
visible change in line profiles).
Apparently, the correction becomes important when most of the radiation is
generated near a black-hole horizon (typically, $r\lta20\,r_g,$
$r_g\equiv1+\sqrt{1-a^2})$ or when observer inclination is large
$(\theta_{\rm obs}\gta75\dg$). The approximation of
straight photon trajectories is unable to reflect lensing effects and we thus
turn to a more sophisticated treatment
by solving the geodesic equation in the Kerr metric.

Fig.~\ref{fig2} illustrates variations of the line profile of a source
in the Keplerian circular motion around a Schwarzschild black
hole $(a=0).$ It is assumed that the source has the form of a spot
radiating isotropically in its local frame. Local emissivity has a
Gaussian frequency profile $(\mbox{FWHM}\approx0.2).$ Local intensity decreases
exponentially with the distance from the centre of the spot.
The characteristic diameter $d$ of the spot is defined by requiring that
intensity at the edge of the spot decreases by a factor $1/e$ in
comparison with the value in the
centre. In our example, $d\approx1\,r_g.$ The
resulting observed profile is not sensitive to the assumptions
above, provided
\beq
2\pi r_{\rm em}\gg d,\quad
\mbox{FWHM}\ll\frac{\nu_{\rm max}-\nu_{\rm min}}{\nu_{\rm max}+\nu_{\rm min}}.
\eeq
The local profile of
the line is centred around $\nu_{\rm em}$ but the observed
profile oscillates around a lower value of frequency
(observed centroid frequency) because of the
gravitational redshift. The maximum frequency corresponds to the source
approaching the observer $(\varphi\approx\frac{3}{2}\pi)$, the
minimum frequency corresponds to the source receding from the
observer $(\varphi\approx\frac{1}{2}\pi)$. The observed count rate follows
this variation strictly only if lensing and time delay are
ignored (cf.\ eq.~[\ref{f}]). On the other hand, once these effect are
take into account properly, the count rate has its maximum
somewhere within the interval $\pi\lta\varphi\lta\frac{3}{2}\pi,$ depending on
$\theta_{\rm obs}$. Two well-separated local maxima of the count rate
can be found in the signal. Lensing effect is responsible for enhancement at
$\varphi\approx0.5$ but the corresponding peak is visible only if
$\theta_{\rm obs}\gta75\dg$ (Fig.~\ref{fig2}a).
Doppler shift produces a peak at $\varphi\approx0.7.$ One can estimate
inclination by fitting the line profiles resolved in frequency and time,
while with frequency integrated light curves such an estimate is
rather uncertain (it can be shown that the phase of the maximum of frequency
integrated signal depends only weakly on $a$ and $\theta_{\rm obs}$; cf.
Fig.~8 of Karas, Vokrouhlickì \& Polnarev 1992).
Since the geodesic equation was solved for each of the rays, the correct
relation between the orbital phase $\phi$ of the source on the disk surface
and the the phase $\varphi$ of the observed signal is automatically
taken into account. In the case of a rotating black
hole, line profiles take a similar course but the frequency range
and phase of the maximum are different, as
illustrated in the next figure.

A relation between the observed frequency and the count rate is further
explored in Fig.~\ref{fig3} where the centroid frequency of the line is given
in terms of the count rate. The variable part of the count rate is
normalized to unit interval.
The centroid frequency is normalized to the local centroid frequency of the
line profile in the frame comoving with the source. It is worth
noticing that the maximum and the minimum centroid energy
($\nu_{\rm c,max},$ $\nu_{\rm c,min})$ and the observed frequency of the
maximum
count rate ($\overline{\nu}$) are specific indicators which relate each of the
curves to the corresponding values of $\theta_{\rm em},$ $r_{\rm em},$ and $a$
(Table~1). One can define the relative difference of the centroid
frequencies:
\beq
\Delta\nu_{\rm c}\equiv\frac{\nu_{\rm c,max}}{\nu_{\rm c,min}}-1
\eeq
which increases with $\theta_{\rm obs}$ increasing and
$r_{\rm em}$ decreasing. Large values of $\Delta\nu_{\rm c}$ $(\gta0.5)$
correspond to relativistic motion of the orbiting source, while small
$\overline{\nu}$ $(<\nu_{\rm c,max})$
indicate the focussing effect to be important.
One can verify from Tab.~1 that, for large inclination, $\overline{\nu}$
decreases with $r_{\rm em}$ increasing, while for small
inclination the dependence is reversed. Parameters
$\Delta\nu_{\rm c}$ and $\overline{\nu}$ supplement the asymmetry parameter
$\cal A$ which was introduced by Gerbal \& Pelat (1981, eq.~[13]) in order to
characterize asymmetry of the time-integrated line profiles.

\bigskip\bigskip\bigskip\par

\noindent
{\large\bf 3.\quad Conclusion}
\bigskip\bigskip\par

The sufficient resolution, both in time and
frequency, is crucial for practical applicability of Doppler tomography
and this is presently the main difficulty of the method
in studying inner regions of accretion disks in AGNs.
In the present contribution we concentrated on properties of radiation
from a single orbiting source, taking into account gravitational lensing
of a Kerr black hole. This is the point where the situation
differs from previous works---otherwise the techniques of image
restoration can be applied in a similar manner as has been done
for tracing, e.g., stellar surfaces or detached binary systems.

Orbital motion of the source around a black hole results in variations
of the observed count rate and increases the efficiency parameter of the
source. This may be a relevant explanation for some extremely variable X-ray
objects from the I~Zw~1
class, and its relevance can be tested by studying line profiles in time.
The method restricts possible values of inclination angle of the
disk and angular momentum of the black hole, and it should thus be taken
into account in planning future X-ray observations.

We assumed that the source of light radiates isotropically in its local
comoving frame, and that it rotates in accordance with the
Kepler law. The assumption about local emissivity does not play a
particular role in our discussion of the line profiles (Fig.\
\ref{fig1}) because we did not consider general
relativistic bending of light trajectories. Fully relativistic treatment
of Figs.\ \ref{fig2}--\ref{fig3} is however affected by this constrain.
Also the rotation law needs to be determined in
accordance with the physical situation (equatorial motion around a Kerr
black hole, in our case). Otherwise, it introduces another free parameter
which must be determined by systematic exploration of the parameter
space and fitting values from Tab.~1 to actual data.

This work has been supported by the grants GACR 205/\-94/\-0504 and
GACR 202/\-96/\-0206 in Czech Republic, and by Wenner-Gren Center Foundation
in Sweden.

\newpage
\parindent=-15pt
{\large\bf References}
\bigskip\par
\parskip=1pt

Abramowicz M. A., Bao G., Lanza A., Zhang X.-H. 1991, A\&A 245, 454

Abramowicz M. A., Lanza A., Spiegel E. A., Szuszkiewicz E. 1992, Nature 356, 41

Albright G. E., Richards M. T. 1996, ApJ 459, L99

Asaoka I. 1989, PASJ 41, 763

Bao G., Hadrava P., {\O}stgaard E. 1994, ApJ 435, 55

Boller Th., Tr\"umper J., Molendi S., Fink H. et al.\ 1993, A\&A 279, 53

Bromley B. C., Chen K., Miller W. A. 1996, ApJ, submitted

Eracleous M., Halpern J. P. 1994, ApJSS 90, 1

Chen K., Halpern J. P., Filippenko A. V. 1989, ApJ 339, 742

Fabian A. 1979, Proc. Roy. Soc. London, Ser. A 366, 449

Fabian A. C., George I. M. 1991, in Iron-Line Diagnostics in X-Ray Sources,
 eds A.~Treves, G.~C. Perola, L.~Stella (Springer-Verlag, Berlin), p.~169

Feigelson E. D., Bradt H., McClintock J., Remillard R. et al.\
 1986, ApJ 302, 337

Forster K., Halpern J. P. 1996, ApJ, to appear

Gerbal D., Pelat D. 1981, A\&A 95, 18

Hadrava P. 1995, A\&AS 114, 393

Kaitchuck R. H., Schlegel E. M., Honeycutt R. K., Horne K. et al.\
 1994, ApJSS 93, 519

Karas V., Lanza A., Vokrouhlick\'y D. 1995, ApJ 440, 108

Karas V., Vokrouhlick\'y D., Polnarev A. 1992, MNRAS 259, 569

Kojima Y. 1991, MNRAS 250, 629

Kwan J., Chen F.-Z., Fang L.-Z., Zheng W., Ge J. 1995, ApJ 440, 628

Laor A. 1991, ApJ 376, 90

Luminet J.-P. 1979, A\&A 75, 228

Marsh T. R., Horne K. 1988, MNRAS 235, 269

Matt G., Perola G. C., Piro L., Stella L. 1992, A\&A 257, 63

Mushotzky R. F., Done C., Pounds K. A. 1993, ARA\&A 31, 717

Papadakis I. E., Lawrence A. 1995, MNRAS 272, 161

Phillips M. M. 1976, ApJ 208, 37

Pineault S., Roeder R. C. 1977, ApJ 213, 548

Rauch K. P., Blandford R. D. 1994, ApJ 421, 46

Rees M. J. 1984, ARA\&A 22, 471

Serlemitsos P. J., Jalota L., Soong Y., Kunieda H. et al.\ 1995, PASJ 47, 105

Simon K. P., Sturm E. 1994, A\&A 281, 286

Sivron R., Caditz D., Tsuruta S. 1996, ApJ, in press

Vogt S. S., Penrod G. D., Hatzes A. P. 1987, ApJ 321, 496

\newpage
\parskip 25pt
\parindent 0pt
{\tt Figure Captions }
\bigskip\bigskip

Fig.~\ref{fig1}. Lower panels: Lines of equal frequency shift
$g$ on an equatorial accretion disk
in Keplerian rotation around a Schwarzschild black hole.
Observer looks from the bottom of the page $(\phi=0)$. Both graphs
are drawn with radial coordinate $x\equiv1-3\,r_g/r$ $(r_g=2,$ $0\leq
x\leq1).$
The whole equatorial plane outside the inner edge of the disk is thus
captured in this illustration; dotted circles corresponds to $x=1$.
Observer inclination is (a) $80\dg$ and (b) $20\dg$.
Upper panels:
Observed count rates (eq.~[\ref{f}]) from a point-like source orbiting at
$r_{\rm em}=3\,r_g$ (solid line) and $r_{\rm em}=8\,r_g$ (dashed line).
The count rate is in arbitrary units; frequency is normalized to
$\nu_{\rm em}$.

Fig.~\ref{fig2}. Time variation of the line profile from a source near a
Schwarzschild black hole;
(a) $r_{\rm em}=14\,r_g,$ $\theta_{\rm obs}=80\dg;$
(b) $r_{\rm em}=14\,r_g,$ $\theta_{\rm obs}=20\dg;$
(c) $r_{\rm em}=4\,r_g,$ $\theta_{\rm obs}=20\dg.$
Count rate is normalized to the maximum value;
frequency is normalized to $\nu_{\rm em}$. The
phase interval corresponding to one revolution of the source is shown
$(0\leq\varphi\leq1)$. See the text for details.

Fig.~\ref{fig3}. Centroid frequency versus count rate. Each loop
characterizes one complete revolution of the source at given
$r_{\rm em}.$ Three different radii of the orbit have been chosen, as
indicated by the line-style: $r_{\rm em}=3\,r_g$ (solid line);
$r_{\rm em}=5\,r_g$ (dashed line); $r_{\rm em}=8\,r_g$ (dot-dashed
line).
Four panels are shown:
(a) $\theta_{\rm obs}=80\dg,$ $a=0;$
(b) $\theta_{\rm obs}=20\dg,$ $a=0;$
(c) $\theta_{\rm obs}=80\dg,$ $a=1;$
(d) $\theta_{\rm obs}=20\dg,$ $a=1.$
Results from this figure are summarized in Tab.~1.

\vspace{1cm}

\pagestyle{empty}


Hardcopy of figures are available upon request from the
authors.
You can also retrieve the complete postscript file of this paper.

\begin{figure}[h]
\caption{V. Karas \& P. Kraus \label{fig1}}
\vspace*{1cm}
\framebox{\hspace*{\hsize}}
\vspace*{1cm}
\end{figure}

\begin{figure}[h]
\caption{V. Karas \& P. Kraus \label{fig2}}
\vspace*{1cm}
\framebox{\hspace*{\hsize}}
\vspace*{1cm}
\end{figure}

\begin{figure}[h]
\caption{V. Karas \& P. Kraus \label{fig3}}
\vspace*{1cm}
\framebox{\hspace*{\hsize}}
\vspace*{1cm}
\end{figure}

\begin{table}[p]
\begin{center}
Table 1. Characteristics of the centroid frequency
\bigskip\par
\begin{tabular}{ccccccccccc}
\multicolumn{2}{c}{}      &
\multicolumn{4}{c}{$a=0$}  &
\multicolumn{1}{c}{}      &
\multicolumn{4}{c}{$a=1$} \\ \cline{3-6} \cline{8-11}
$\theta_{\rm obs}$ [deg] & $r_{\rm em}/r_g$
   & $\nu_{\rm c,max}$ & $\nu_{\rm c,min}$ & $\Delta\nu_{\rm c}$
    & $\overline{\nu}$ &
   & $\nu_{\rm c,max}$ & $\nu_{\rm c,min}$ & $\Delta\nu_{\rm c}$
    & $\overline{\nu}$
     \rule[-1.5ex]{0ex}{4ex} \\ \hline \hline
80 & 3 & 1.32 & 0.53 & 1.52 & 1.13 & & 1.31 & 0.51 & 1.55
 & 1.19\rule[0ex]{0ex}{3ex} \\
20 & 3 & 0.84 & 0.60 & 0.39 & 0.83 & & 0.68 & 0.52 & 0.29 & 0.67 \\
80 & 5 & 1.26 & 0.62 & 1.05 & 1.00 & & 1.31 & 0.51 & 1.55 & 1.08 \\
20 & 5 & 0.95 & 0.74 & 0.28 & 0.95 & & 0.83 & 0.59 & 0.41 & 0.83 \\
80 & 8 & 1.23 & 0.71 & 0.74 & 0.99 & & 1.27 & 0.58 & 1.20 & 0.93 \\
20 & 8 & 0.98 & 0.82 & 0.18 & 0.98 & & 0.92 & 0.71 & 0.28
 & 0.92\rule[0ex]{0ex}{-1.5ex} \\ \hline
\end{tabular}
\end{center}
\end{table}

\end{document}